# The Impact of COVID-19 on Stock Market Volatility in Pakistan

Dr. Ateeb Akhter Shah Syed[1]
Dr. Kaneez Fatima[2]

**Abstract**

This paper examines the impact of coronavirus (COVID-19) on stock market volatility (SMV) in Pakistan by controlling the effect of exchange rate, interest rate and government/central bank interventions to combat the pandemic. We used the vector autoregressive (VAR) model over a sample period ranging from February 25, 2020 to December 7, 2020. We find that a shock to total daily coronavirus cases in Pakistan lead to a significant increase in SMV. This result is aligned with a vast literature on pandemics and investors' uncertainty and remains robust to several robustness checks applied in our analysis.

[1] Corresponding Author: Research Department, State Bank of Pakistan (SBP), 8th Floor, Main Building, I.I.Chundrigar Road, Karachi, Sindh, Pakistan, 74000. Email: ateeb@fulbrightmail.org, ORCID ID: 0000-0003-3872-9609

[2] Institute of Management Sciences, University of Balochistan

**Introduction**

This paper aims to investigate the impact of COVID-19 on SMV in Pakistan by considering the other major factors that influence the performance of stock markets such as interest rate and exchange rate. Most importantly, we consider the measures taken by government in early response of pandemic such as closure of business activities and implementing smart lock down later by relaxing the restrictions. Several economic measures such as fiscal stimulus and easing of monetary policy to mitigate the slowdown in economic activities are also taken into account. The Pakistani stock market posted record losses due to jerks to the sentiments of investors facing the lowest six-year intra-day value, that is, a reduction by 28 percent this year Pakistan Institute of Development Economics (2020).

To mitigate the impact of the pandemic the government of Pakistan approved a fiscal stimulus (a package of Rs. 1.2 trillion and Supplementary Grant of Rs.100 billion for the "Residual/Emergency Relief Fund" in relation to provision of funds for mitigating the effect of COVID-19 for the impacted population). Similarly, the SBP took several measures to combat the damage to the economy caused by the pandemic. These measures include reduction of policy rate by 625 basis point, extension of time for settlement of foreign currency loans, and several subsidized financing schemes.

During pandemics, wars, natural disasters and financial crises, the level of uncertainty in the markets is exceptionally high and so is the level of risk aversion among investors [Baker et al. (2020); Eichenbaum, Rebelo, & Trabandt ( 2020)]. Historically, the stock markets had been responsive to epidemics Wang, Yang, & Chen (2013). The stock markets around the globe reacted forcefully to the COVID-19 which was unprecedented in many stock markets (Baker et al., 2020). The literature on impact of -19 on stock market is limited but growing quickly. Number of studies investigate the impact of pandemic on the performance of different stock markets [Mzoughi et. al (2020); Onali (2020); Zaremba et. al (2020)]. These studies provide an evidence of significant increase in SMV in response to COVID-19.

The literature on response of stock market in Pakistan is however, limited. Waheed et. al (2020) tested the impact of COVID-19 on Karachi Stock Exchange (KSE). They provide an evidence of positive impact on performance of KSE attributing it to timely response of Government. However, they do not control for the measures taken by the government and other fundamental economic variables which affect the performance of stock market. Yar (2020) attempted to determine the performance of stock market in response to daily positive cases of COVID-19, fatalities and recoveries. They find insignificant impact of positive cases and fatalities on the stock market. This evidence in contradiction to what is observed globally, may be due to overlooking the other important factors that influence the markets.

The response of SMV in Pakistan is not investigated in literature. Therefore, this paper contributes to the literature by investigate the response of volatility of the KSE to COVID-19.

**Material and Method**

The variables used in this study are: total daily coronavirus cases (C) as a measure of COVID-19, weighted average overnight repo rate (R) (current policy rate of SBP), Pakistani Rupee to US Dollar nominal exchange rate (E) as a measure of exchange rate (US Dollar is the reserve currency of Pakistan), stringency index (S) and SMV. These variables are added to the model both on account of their availability. A dummy variable (D) is added to account for fiscal stimulus and monetary easing during the pandemic.[3] The data is daily from February 25, 2020 to December 7, 2020. The data on C and S is available for all days of the week whereas data on other variables is on working days only. To match the data with C, the value of last working day of the other variables is carried forward for the missing days (weekends and holidays). Data of C and S are taken from www.ourworldindata.org, data on KSE general index of all shares price index (SM) is taken from the Bloomberg and data on all other variables is taken from SBP.

Volatility of the stock market is measured using the conditional variance of the first difference of SM (DSM). The mean model is an AR(1,6) specified using the Box-Jenkins (1976) methodology and the SMV is obtained from a generalized autoregressive conditional heteroskedastic (GARCH) model Bollerslev (1986). The adequacy of the mean model and the GARCH(2,1) model is ensured by testing the residuals from mean model and the squared standardized residuals from the GARCH(2,1) model using the Ljung-Box (1978) test, respectively (results of volatility measurement not reported to conserve space but available upon request).

Introduced by Sims (1980), we use the VAR for estimation. All the variables except R, are in log-levels. All the variables are tested for stationarity using a series of Dickey-Fuller (1981) tests and found stationary in log-levels/levels. Therefore, we estimate the model in log-levels/levels.[4] A series of robustness check exercises[5] have been conducted to ensure robustness of our findings.

**Results**

Figure 1 shows the response[6] of SMV to a shock in C from the base model, shock to C produces a significant positive impact on SMV for the first 5 forecast horizons and is insignificant thereafter. This result in line with Onali (2020) and Zaremba et. al (2020) among others and indicates that during pandemic investors uncertainty rises. However, given experiences such as the great depression and especially the great recession in the information kit, the SBP and the government responded with policy measures in a timely manner. This may explain why the impact of COVID-19 on investors' uncertainty is not as big and prolonged as it could have been otherwise.

---

[3] The dummy variable D takes on value of "1" at March 17, 2020 and April 1, 2020 and is 0 otherwise. It is because the SBP started to lower the policy rate from March 17, 2020 and an emergency cash program was launched by the government on April 1, 2020.

[4] Our baseline VAR model contains variables C, R, E, SMV and D and it is estimated using 20 lags. The estimation period is from March 17, 2020 to December 7, 2020. The lag length was increased until we found white noise residuals from each VAR equation. Ljung-Box (1978) test showed that the residuals from the VAR equations are white noise. The ordering for the base model is C, R, E, SMV. This ordering is consistent with the hypothesis we are testing: C is a health crisis driven by the pandemic and is assumed to be exogenous. R is placed next to control for the monetary policy response to the pandemic. A change in R leads to a change in E, this places E after R. Finally, as SMV reflects developments in the economy and especially in the R and E quickly; therefore, it is placed last in the ordering.

[5] We performed several robustness checks: First, we added the stringency index as a measure of government intervention (closure of business activities, schools, universities, and other social distancing measures). Second, we estimated the model by removing the exchange rate variable. We also estimated a bi-variate VAR (C and SMV) with reverse ordering. As china and US are the top two trading partners of Pakistan. Hence, finally we also used Pakistani Rupee to Chinese Yuan (CNY) as a measure of exchange rate. All these robustness exercises revealed that the results qualitatively remain the same.

[6] The confidence intervals for the IRFs are computed via ten thousand Monte Carlo draws. For each IRF, we report the bootstrapped confidence bands indicate the two-standard deviation of the draws.

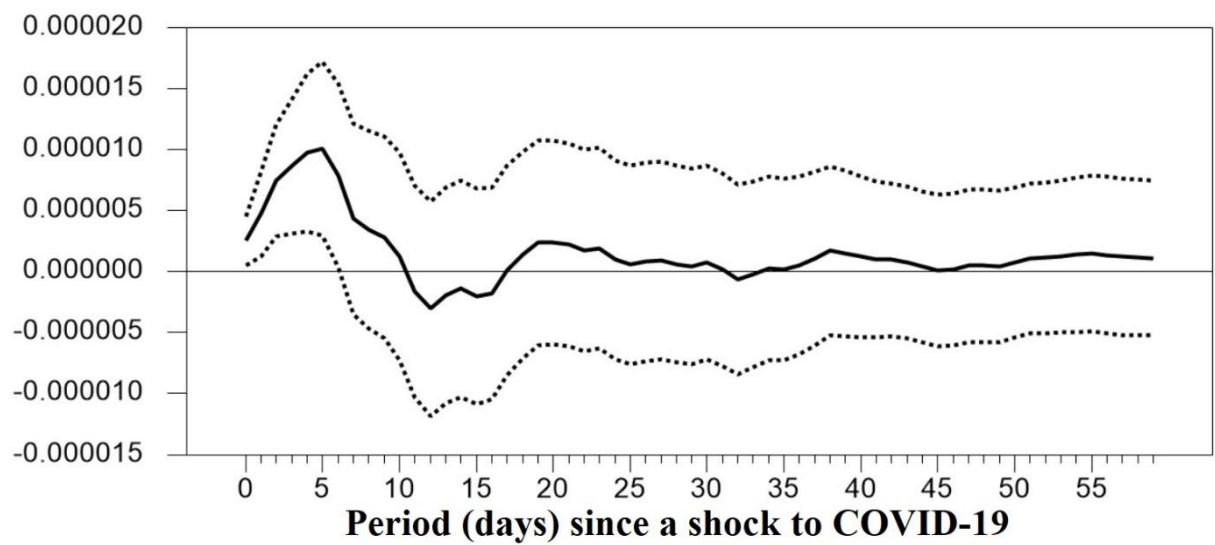

**Fig.** 1 Impact of C on SMV (base model). The figure show VAR-estimated impulse response functions (60 days) for SMV to COVID-19 innovations (solid black line) with confidence bands (dotted black lines).

**Conclusion**:

The stock markets unprecedently responded to the COVID-19 around the globe by increased volatility and decreased returns. However, preliminary evidence in literature for Pakistan is not inline with the evidence on global stock markets ((Waheed et al., 2020) and (Yar, 2020a)). In this paper we investigated the impact of COVID-19 on SMV using a daily data over the sample period February 25, 2020 - December 7, 2020. We control the effect of economic variables (E & R), and policy measures of government to mitigate the losses posed by the pandemic. We used a VAR model to estimate the impact of COVID-19 on SMV.

Our baseline model shows that SM demonstrates a significant rise in its volatility in response to a COVID-19 shock. These results are robust to different measures of E and specifications of the model. These results are consistent with the results reported in the recent literature. However, the response of volatility is not very high and became insignificant after five days which may be attributed to measures taken by policy makers to manage the damaging confidence and risk averse behavior of investors. Moreover, Pakistan was not hit as hard by COVID-19 as many other developed economies were.